\documentclass[
aps,
article,
longbibliography,
superscriptaddress,
 amsmath,amssymb,
twocolumn,
]{revtex4-1}

\usepackage{graphicx}
\usepackage{dcolumn}
\usepackage{bm}
\usepackage{upgreek}
\usepackage{color}
\usepackage{hyperref}

\newcommand{\dg}{\dagger}

\newcommand{\fref}[1]{Fig.~\ref{#1}}
\newcommand{\puoli}{\frac{1}{2}}

\newcommand{\vari}[1]{{#1}}

\definecolor{gold}{RGB}{215,155,0}
\definecolor{blue}{RGB}{0,0,255}
\definecolor{red}{RGB}{255,0,0}
\definecolor{darkgreen}{RGB}{20,150,10}
\definecolor{darkblue}{RGB}{10,10,150}
\definecolor{orange}{RGB}{200,100,0}
\definecolor{lightblue}{RGB}{50,150,230}

\usepackage[normalem]{ulem}

\newcommand{\CHANGE}[1]{{#1}}

\begin{document}

\title{Quantum-mechanics free subsystem with mechanical oscillators}

\author{Laure Mercier de L\'epinay}
\affiliation{QTF Centre of Excellence, Department of Applied Physics, Aalto University, FI-00076 Aalto, Finland}
\author{Caspar F. Ockeloen-Korppi\footnote{Present address: IQM Finland Oy, Keilaranta 19 02150 Espoo, Finland}}
\affiliation{QTF Centre of Excellence, Department of Applied Physics, Aalto University, FI-00076 Aalto, Finland}

\author{Matthew J. Woolley}
\affiliation{School of Engineering and Information Technology, UNSW Canberra, ACT, 2600, Australia}

\author{Mika A.~Sillanp\"a\"a}
\affiliation{QTF Centre of Excellence, Department of Applied Physics, Aalto University, FI-00076 Aalto, Finland}

\begin{abstract}
Quantum mechanics sets a limit for the precision of continuous measurement of the position of an oscillator. Here we show how it is possible to measure an oscillator without quantum backaction of the measurement by constructing one effective oscillator from two physical oscillators. We realize such a quantum-mechanics free subsystem using two micromechanical oscillators, and show the measurements of two collective quadratures while evading the quantum backaction by $8$ decibels on both of them, obtaining a total noise within a factor two of the full quantum limit. This facilitates detection of weak forces and the generation and measurement of nonclassical motional states of the oscillators. Moreover, we directly verify the quantum entanglement of the two oscillators by measuring the Duan quantity $1.4$ decibels below the separability bound.
\end{abstract}


\maketitle

Measuring a quantum-mechanical system without disturbing it is not possible. 
Consider an oscillator of angular frequency $\omega_0$ and dimensionless position $x(t)$ and momentum $p(t)$. The measurement of $x(t)$ causes $p(t)$ to suffer a disturbance called quantum backaction (QBA). The disturbance of momentum dynamically leads to a disturbance of position and therefore a fundamental limit on continuous position measurement. The oscillator's position can be written $x(t) = X\sin (\omega_0 t) + P \cos (\omega_0 t)$ where, in a quantum-mechanical framework, the  quadratures of the motion $X$ and $P$ are non-commuting conjugate observables that cannot be known simultaneously with arbitrarily high precision: if $X$ is measured then $P$ is subject to backaction, and vice versa.

In a backaction evading (BAE) \vari{measurement strategy} \cite{BraginskyQND} a probe couples to only one quadrature of the oscillator's motion, say $X$. The backaction associated with this measurement disturbs the $P$ quadrature, but the disturbance is not fed back to the measured $X$ quadrature. Therefore, the $X$ quadrature can be measured without any fundamental limit, at the expense of lost information on the $P$ quadrature \cite{Schwab2014QND,Schwab2016squ3db,2BAE,Polzik2017,Kippenberg2019BAE}. 

\begin{figure}[t]
  \begin{center}
    {\includegraphics[width=0.9\columnwidth]{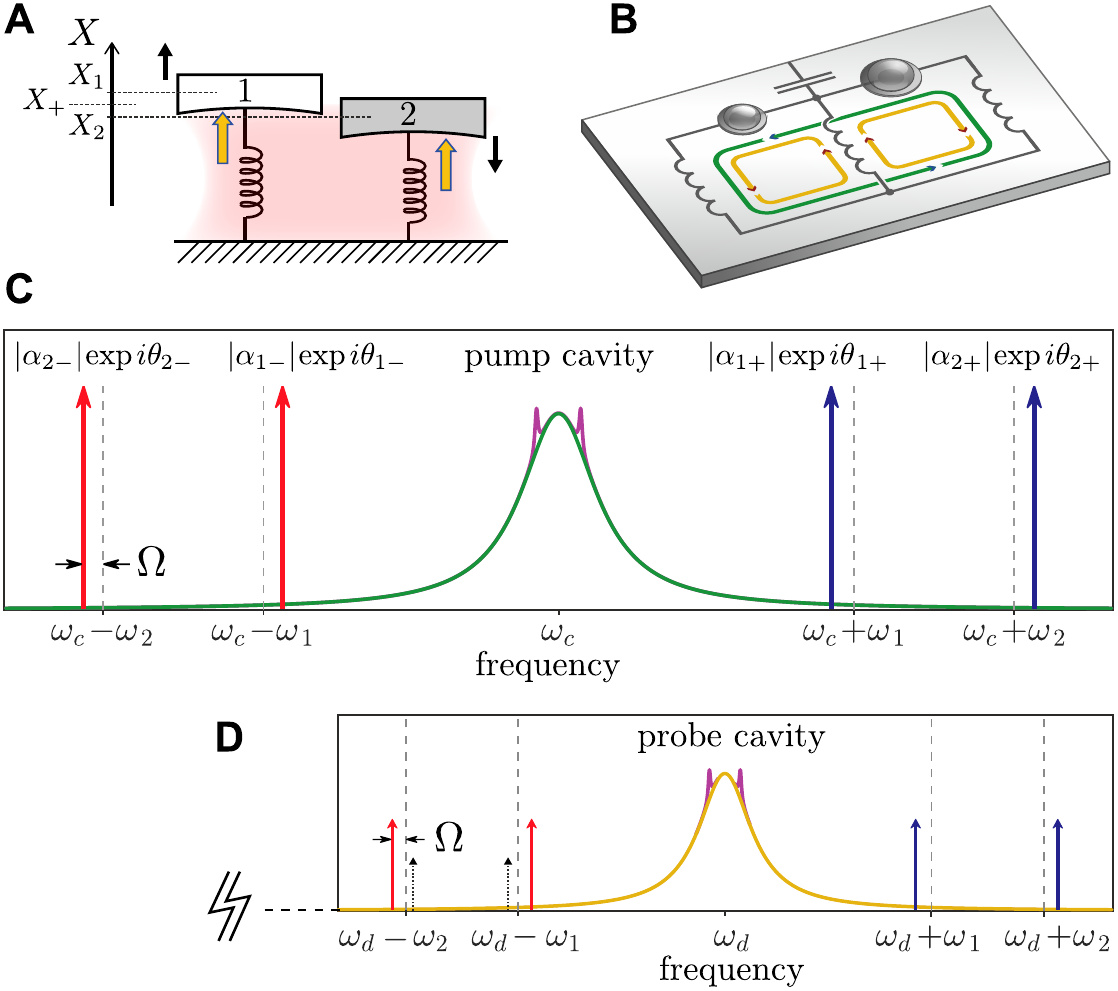}} 
    \caption{\textit{Four-tone quantum backaction evading measurement.} \textbf{(A)} Illustration of a quantum-mechanics-free subsystem (QMFS) constructed from an effective positive-mass oscillator (1, white) and an effective negative-mass oscillator (2, shaded) coupled to a cavity. Under the same force applied on both oscillators (wide arrows), e.g.~the fluctuating radiation-pressure force, both momenta $P_{1,2}$ are changed by the same amount, but oscillator 2 is displaced in the opposite direction to oscillator 1 due to its negative mass. The sum momentum $P_+ = (P_1 + P_2)/\sqrt{2}$ is therefore dynamically coupled to the difference position $X_- = (X_1 - X_2)/\sqrt{2}$, just as per the two quadratures of a single oscillator. However, unlike a standard oscillator's quadratures, these commute: as a result, $\{X_-, P_+\}$ span a QMFS, and similarly for $\{X_+, P_-\}$. \textbf{(B)} Physical realization with two Al drumhead oscillators coupled to two microwave resonances in a superconducting microcircuit. \CHANGE{The arrows denote currents in the two modes, see Fig.~S1.} \textbf{(C)} An optomechanical cavity (frequency $\omega_c$) is pumped by four coherent tones at frequencies close to the red and blue sidebands of two mechanical oscillators having frequencies $\omega_1$ and $\omega_2$, detuned by $\Omega$. \textbf{(D)} The oscillators are also coupled to an auxiliary (probe) cavity that is used for tomographic monitoring with four probe tones, and for sideband cooling (dashed black arrows). \CHANGE{In \textbf{(C)}, \textbf{(D)}, the set of small peaks centered on cavity resonance denote the sideband processes, spaced by $2\Omega$.}}
    \label{fig:scheme}
 \end{center}
\end{figure}

Remarkably, by constructing an effective oscillator from two physical oscillators one can monitor both quadratures of an oscillator without any fundamental limit. This involves pairing effective positive and negative mass oscillators as illustrated in \fref{fig:scheme}A \cite{Caves2010BAE,Caves2012BAE,Meystre2013bae,Vitali2016QMFS}, such that QBA is coherently cancelled for two of their collective quadratures. 
This approach creates a quantum-mechanics-free subsystem (QMFS) 
\cite{Caves2010BAE,Caves2012BAE}. This possibility is particularly crucial in several ultra-sensitive measurements beating conventional quantum limits and studies of nonclassical states of motion \cite{Entanglement}. 


The principle of the scheme has been demonstrated in experiments using optomechanics and spin-mechanics hybrid systems \cite{Heidmann2007BAE,2BAE,Polzik2017}. 
Here, we present direct evidence of the absence of QBA on the collective dynamics of a pair of mechanical oscillators. 
Our system consists of two aluminum membrane micromechanical oscillators \cite{Teufel2011a} \CHANGE{optomechanically coupled} to two microwave cavity modes (\fref{fig:scheme}B).
One of the modes, called the pump cavity mode (\fref{fig:scheme}C), is characterized by the frequency $\omega_c/2\pi$ and damping rate $\kappa/2\pi$ and described by the annihilation and creation operators $a$, $a^\dg$. The mechanical oscillators, labeled 1 and 2, have frequencies $\omega_1/2\pi$ and $\omega_2/2\pi$, and damping rates $\gamma_1/2\pi$ and $\gamma_2/2\pi$. The average damping rate is $\gamma  \equiv (\gamma_1 +  \gamma_2)/2$. The creation and annihilation operators for phonons are denoted $b_j$, $b_j^\dg$, with $j=1,2$. Additionally, both oscillators are coupled to a probe cavity (\fref{fig:scheme}D) of frequency $\omega_d/2\pi$ and damping rate $\kappa_d/2\pi$.

In order to create the effective interaction that realizes the QMFS, we drive the cavity with four coherent pump tones \cite{ClerkEnt2014}.
The angular frequencies of the tones are $\omega_{1\pm} = \omega_c \pm (\omega_{1} - \Omega)$ and $\omega_{2\pm} = \omega_c \pm (\omega_2 + \Omega)$ 
(Fig.~\ref{fig:scheme}C). Here, $\Omega > 0$ is a detuning from respective motional sidebands.
We describe the system in a frame \CHANGE{oscillating at} $\omega_c$ for the cavity field, and $(\omega_{1} - \Omega)$ and $(\omega_2 + \Omega)$ for the two oscillators.
In this frame, the mechanical quadratures are $X_j=(b^\dg_j+b_j)/\sqrt{2}$ and $P_j=i(b^\dg_j-b_j)/\sqrt{2}$ for $j=1,2$,
and collective mechanical quadratures are $X_\pm = \frac{1}{\sqrt{2}} \left(X_1 \pm X_2\right)$, and $P_\pm = \frac{1}{\sqrt{2}} \left(P_1 \pm P_2\right)$. 

With strong driving tones, the radiation-pressure interaction \CHANGE{is linearized} \cite{OptoReview2014}. Each pump tone gives rise to an effective complex-valued optomechanical coupling strength 
$|G_{j\pm}| e^{i\theta_{j\pm}}$ associated with the tone of frequency $\omega_{j\pm}/2\pi$. 
Initially, we choose all amplitudes $|G_{j\pm}| \equiv G$ to be equal, and define the cooperativity $C = 4 G^2 /(\kappa \gamma)$. 
We assume $\omega_j \gg \kappa$ and $|\omega_1-\omega_2| \gg \kappa$,
and write the  Hamiltonian as the sum of an uncoupled part and a coupling term; $H=H_0 + H_c$. The uncoupled part $H_0/\hbar = \Omega/2   \left(X_1^2 + P_1^2 - X_2^2 - P_2^2 \right)$ attributes a negative mass to oscillator 2 as required to generate a QMFS (see \fref{fig:scheme}A). In terms of collective quadratures (see Eq.~S4)
\begin{equation}
\begin{split}\label{eq:Heff}
\frac{H_0}{\hbar} & 
 = \Omega  \left(X_+ X_- +  P_+ P_- \right), \\[2pt]
\frac{H_c}{\hbar} & = \frac{G}{2}a \left( A_- X_- + A_+ X_+ + B_- P_- + B_+ P_+\right)  +  \rm{h.c.}
\end{split}
\end{equation}
\CHANGE{The coefficients $A_\pm$ and $B_\pm$ are  functions of pump tones' phases.}
Evolution under $H_0$ couples only pairs of commuting mechanical quadratures, so that each subsystem $\{X_-, P_+\}$ and $\{X_+, P_-\}$ \CHANGE{can constitute a QMFS \cite{Caves2012BAE}.}

\begin{figure*}[t]
  \begin{center}
   {\includegraphics[width=0.9\textwidth]{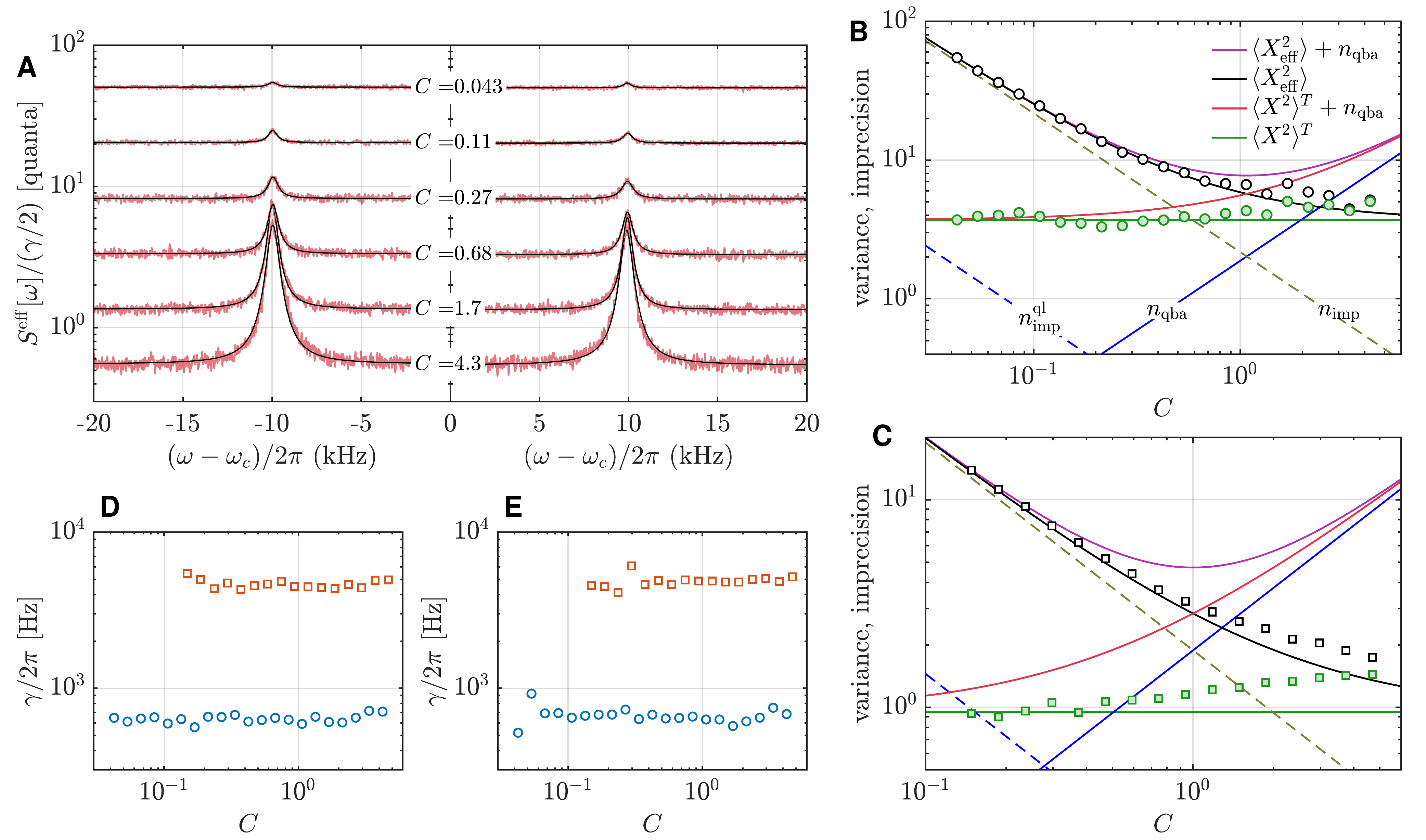} }
    \caption{\textit{Imprecision close to the full quantum limit via two-mode BAE monitoring.} \textbf{(A)} The pump cavity output spectrum when measuring the $X_+$ quadrature shows two peaks whose signal-to-noise level improves as the cooperativity is increased. \textbf{(B)} Mechanical noise of the measurement in (A) showing different contributions. Green solid circles: $\langle X^2\rangle$; black open circles: $\langle X^2_{\rm{eff}} \rangle$. The theoretical lines are labeled. $n_{\rm{imp}}^{\rm{ql}}$ indicates the imprecision with $n_{\rm{amp}} = 0$. \textbf{(C)} As (B), but with stronger sideband cooling. The symbols and lines with a given coloring correspond to those in (B). \textbf{(D, E)} The linewidths corresponding to the left and right peaks in (A) (circles), and similarly for the measurement in (C) (rectangles). Pump detuning is $\Omega/2\pi = 10$ kHz in (A), (B), and $\Omega/2\pi = 200$ kHz in (C).}
    \label{fig:baedata}
 \end{center}
\end{figure*}

For certain combinations of $\theta_{j\pm}$ (Fig.~S6), a BAE measurement of any particular quadrature can be achieved. For example,  $\theta_{1-} = \theta_{2+} = 0$ and $\theta_{1+} = \theta_{2-} \equiv \phi$ realizes 
\begin{equation}
\label{eq:baepm}
\frac{H_c}{\hbar} =  2 \sqrt{2}\,G\,  X_c^\phi \left( X_+ \cos\frac{\phi}{2} + P_- \sin\frac{\phi}{2}\right) \,,
\end{equation}
which couples any linear combination $X_+^\phi \equiv X_+ \cos\frac{\phi}{2} + P_- \sin\frac{\phi}{2}$ of \CHANGE{quadratures} $X_+$ and $P_-$ to a certain quadrature of the cavity field $X_c^\phi= \left(a\, e^{-i\frac{\phi}{2}} + a^\dagger\, e^{+i\frac{\phi}{2}}\right)/\sqrt{2}$. 
The cavity thus measures a given mechanical quadrature, e.g.~$X_+$ for $\phi=0$. In this case there is backaction on the conjugate quadrature $P_+$ but, since the evolution of the subsystem  $\{X_+, P_-\}$ is independent from $P_+$, the disturbance never leaks back to this subsystem which remains an isolated QMFS.

Another case is with $\theta_{1-} = \theta_{2-} = 0$ and $\theta_{1+} = \theta_{2+} \equiv \theta$, which measures an arbitrary linear combination of $X_+$ or $P_+$ depending on $\theta$. 
Similarly, any linear combination of $X_\pm$ and $P_\pm$ can be  measured with an appropriate choice of phases \CHANGE{by detecting the field leaking out of the cavity, on which is imprinted the information on the mechanical quadrature  \cite{WoolleyBAE,2BAE}.}

\CHANGE{Thermal and quantum fluctuations determine the different noise contributions of  position measurement.}
The thermal occupation number of a single oscillator $j$ in equilibrium with a bath of temperature $T$ is $n_j^T = \left[\exp(\hbar \omega_j /k_B T )-1\right]^{-1}$. \CHANGE{The fluctuations of the collective quadratures, including the zero-point fluctuations, are described via their respective variances, e.g.,~ $\langle X_+^2\rangle$, $\langle P_-^2\rangle$} for $X_+$, $P_-$. For simplicity, we do not explicitly write the quadrature label for the measured quadratures. If the two oscillators are in a thermal state, each collective quadrature's 
\CHANGE{variance is \cite{2BAE} $\langle X^2\rangle = \langle X^2\rangle^T$, where $\langle X^2\rangle^T = \puoli \left( n^T_1 + n^T_2 \right) + \puoli$. }

In spite of QBA associated \CHANGE{with} the shot noise of the pump tones \cite{RegalShot,StamperKurn2014SQL,Teufel2016ShotN}, the \CHANGE{variances of the measured quadrature and its pair in the QMFS are still given by $\langle X^2\rangle=\langle X^2\rangle^T$} thanks to BAE.
An additional classical contribution (technical heating), however,
makes  $\langle X^2\rangle^T$ weakly power-dependent. The conjugate quadratures each receive QBA $n_{\rm qba}=2C$ as well as a small classical contribution
$n_{\rm cba} = 4C n_c^T$ \CHANGE{due to cavity thermal occupation $n_c^T$} \cite{WoolleyBAE,2BAE}, such that the \CHANGE{variances of the quadratures orthogonal to the QMFS are $\langle X^2\rangle =  \langle X^2\rangle^T + n_{\rm qba}+ n_{\rm cba}$}.

%
\CHANGE{The measurement also suffers} from imprecision noise that can be written (\CHANGE{see} Eqs.~S32, S33) 
as an equivalent collective quadrature \CHANGE{occupation} $n_{\rm imp}\simeq \frac{1}{8C} \left(n_{\rm amp}+ \puoli \right)$,
here dominated by the number of noise quanta $n_{\rm amp}$ added by the amplifier. In the case of a continuous non-BAE measurement, the trade-off between $n_{\rm imp}$ and $n_{\rm qba}$ defines the standard quantum limit (SQL) \cite{GirvinReview},
where the added noise equals the zero-point motion noise. The SQL has been approached \cite{Lehnert2009SQL,Kippenberg2010SQL} and recently \CHANGE{surpassed} in an off-resonant case 
\cite{Schliesser2018sql}. 
We refer to the resonant-case SQL as the ``full quantum limit'' \CHANGE{(FQL)}. While thermal noise represents a severe obstacle to reaching the SQL, approaching the FQL remains even much more challenging.
In the BAE case, 
SQL and FQL can be beaten for the two concerned quadratures simultaneously, provided that their thermal noise does not dominate.

The output spectrum $S_{{\rm out} }[\omega ]$ from the pump cavity consists of two peaks located at $\pm \Omega$ \CHANGE{with respect to the cavity resonance frequency}
(\fref{fig:scheme}C). With properly chosen pump phases, these peaks faithfully reproduce the spectrum of the measured mechanical quadrature.
After amplification, the recorded spectrum sits on top of a microwave noise background $\left(n_{\rm{amp}} + 1/2 \right)$, which contributes to an effective equivalent mechanical occupation in the inferred mechanical spectrum $S^{\rm eff}[\omega]$ (Eq.~S33). The effective \CHANGE{mechanical variance} can then be determined as \CHANGE{$\langle X^2_{\rm{eff}} \rangle \equiv \gamma S^{\rm eff}[\pm \Omega ]/2 = \langle X^2\rangle^T +  n_{\rm{imp}}$}. 

The measurements are carried out in a dry dilution refrigerator with a base temperature of 8 mK, where we find the mechanical oscillators to have the equilibrium thermal phonon numbers $\CHANGE{n_1^{T,0}} \simeq 32$ and $\CHANGE{n_2^{T,0}} \simeq 24$ which give $\CHANGE{\langle X^2\rangle^{T,0}} \simeq 28$ corresponding to an effective bath temperature of $10.5\,\rm mK$.
The pump cavity has a frequency $\omega_c/2\pi \simeq 4.98$ GHz, damping rate $\kappa/2\pi \simeq 1.58$ MHz divided into internal $\kappa_I/2\pi \simeq 130$ kHz and external $\kappa_E/2\pi \simeq 1.45$ MHz damping. The mechanical frequencies are $\omega_1/2\pi \simeq 6.692$ MHz and $\omega_2/2\pi \simeq 9.032$ MHz, and intrinsic damping rates are $\gamma^0_1/2\pi \simeq 55$ Hz and $\gamma^0_2/2\pi \simeq 84$ Hz. The probe cavity frequency is $\omega_d/2\pi \simeq 6.62$ GHz, and the damping rates are $\kappa_d/2\pi \simeq 1.17$ MHz and $\kappa^d_I/2\pi \simeq 350$ kHz and $\kappa^d_E/2\pi \simeq 820$ kHz. 


The microwave tones are generated from separate phase-locked oscillators (Figs.~S2,S3).
We calibrate the cooperativity of each red-detuned tone individually through its sideband cooling effect. Single-mode BAE measurements \cite{Schwab2014QND} are used to match the cooperativities of the blue-detuned tones. Additionally, we apply two cooling tones (\fref{fig:scheme}D) to the probe cavity in order to independently sideband-cool both oscillators. This broadens the mechanical linewidths up to $\gamma_1 \simeq \gamma_2 = \gamma \gg \gamma_1^0, \gamma_2^0$ \CHANGE{and reduces the variance $\langle X^2\rangle^T$ far below the value $\langle X^2\rangle^{T,0}$ reached with cryogenic cooling alone.}

To demonstrate BAE, 
all pump phases are chosen equal to zero to measure $X_+$. In \fref{fig:baedata}A, we display the measured noise spectra when \CHANGE{initially $\langle X^2\rangle^T \simeq 3.2$} ($\gamma/2\pi \simeq 630$ Hz). From the spectra, we extract the effective imprecision noise $n_{\rm imp}$ as well as the \CHANGE{mechanical variance $\langle X^2\rangle$}. The latter is nearly independent of the measurement strength as seen in \fref{fig:baedata}B. At large cooperativities, it is clear that both $\langle X^2_{\rm{eff}} \rangle$ and $\langle X^2\rangle$ stay well below 
$n_{\rm qba}$, showing a nearly ideal BAE \CHANGE{and thus the existence of a QMFS}. We then intensify the sideband cooling down to \CHANGE{$\langle X^2\rangle^T \simeq 1.0$} ($\gamma/2\pi \simeq 4.6$ kHz), obtaining the \CHANGE{noise} displayed in \fref{fig:baedata}C. All measured noise figures are smaller, $\langle X^2_{\rm{eff}} \rangle$ falling 7 dB below the QBA level at high cooperativity and $\langle X^2\rangle$ reaching $8\,\rm dB$ below. \CHANGE{The effective noise is less than a factor of two from FQL ($\langle X^2_{\rm{eff}} \rangle = 1$). This exceeds the best values reported thus far, achieved in cold-atom optomechanics \cite{StamperKurn2014SQL}. Reaching the FQL is here precluded by thermal noise, which is enhanced by technical heating of the oscillators \cite{2BAE,Entanglement}.} The linewidth of the peaks is not affected by the measurement (\fref{fig:baedata}D,E).

\begin{figure}[h]
  \begin{center}
   {\includegraphics[width=0.99\columnwidth]{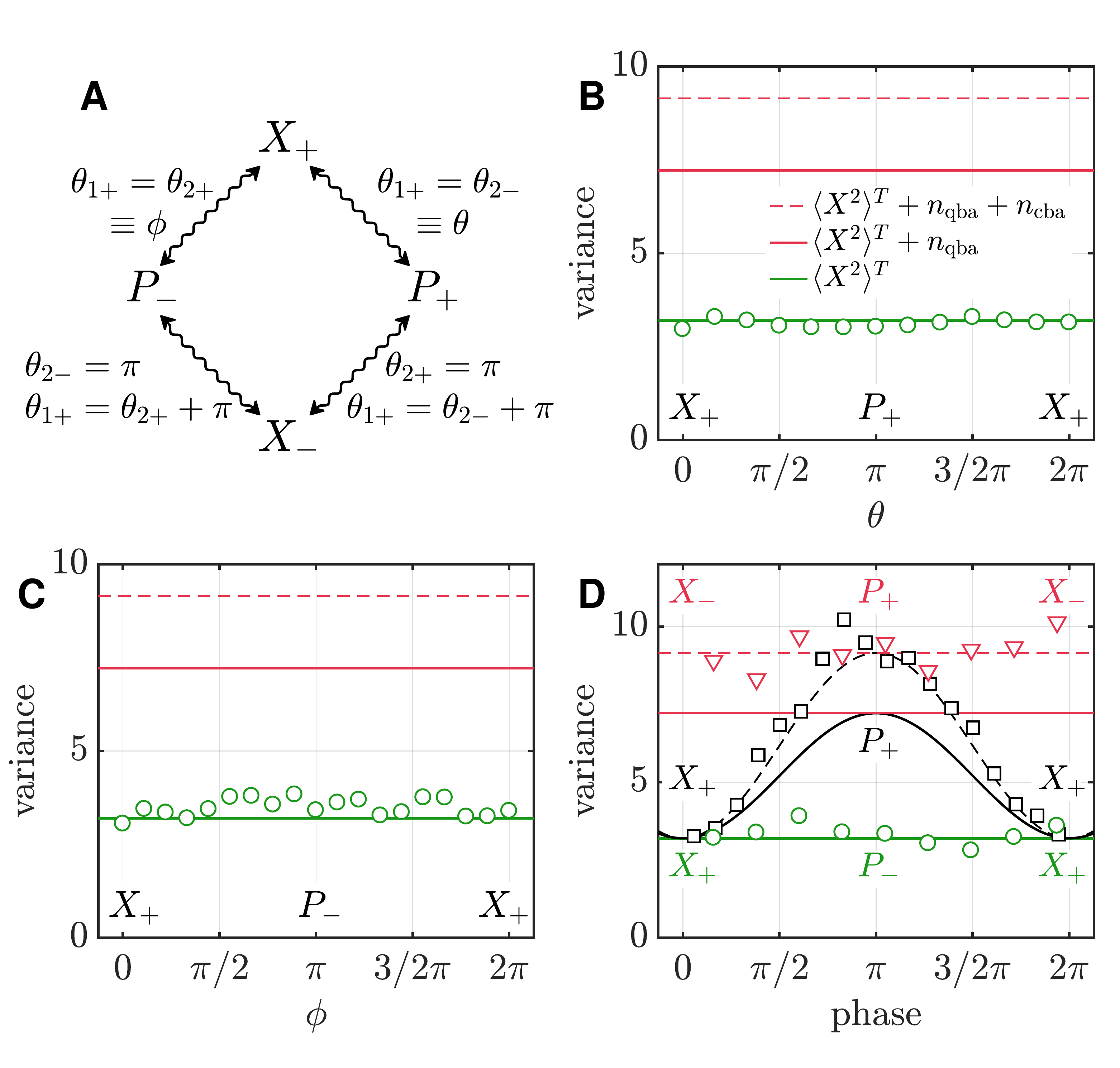}} 
    \caption{\textit{Moving within the quantum-mechanics free subsystem.} \textbf{(A)} Simultaneously shifting two phases  in four-tone BAE as indicated allows the measurement of particular linear combinations of the collective quadratures. This holds for the pumping as well as for tomographic probing. The other unspecified phases in each quadrant of the figure equal zero. \textbf{(B)} Pump cavity signal when shifting between $X_+$ and $P_+$. \CHANGE{The circles denote the measured $\langle X^2\rangle$.} The theoretical lines depict the ideal QBA model, and they are colored similarly to the corresponding data points.  \textbf{(C)} As (B), but shifting from $X_+$ to $P_-$. \textbf{(D)} Probe cavity signal while the pump cavity exerts a strong BAE measurement to $X_+$. The phase sweep is executed as follows: $\phi$ (circles),  $\theta$ (rectangles), or along the route $P_+ \Longleftrightarrow X_-$ in (A) (triangles). These sweeps correspond to moving between different collective quadratures as labeled. The black solid line is the prediction including QBA only, and the black dashed line includes also $n_{\rm cba}$. In (D), the detuning $\Omega/2\pi = 200$ kHz, while in the rest $\Omega/2\pi = 10$ kHz. The pump cooperativity is $C \simeq 2.1$, \CHANGE{and  $\langle X^2 \rangle^T \simeq 3.2$ and $\gamma/2\pi \simeq 630$ Hz}.}
    \label{fig:qmfs}
 \end{center}
\end{figure}

To demonstrate the ability to rotate the QMFS, we now change two phases synchronously.
This operation  is represented in \fref{fig:qmfs}A. In \fref{fig:qmfs}B,C we show the measured \CHANGE{noise} while sweeping the quadrature probed by the probe cavity between $X_+$ and $P_+$, or $X_+$ and $P_-$, respectively. 
Since QBA is always directed to an unmeasured quadrature, the measured \CHANGE{fluctuations} remain at the same backaction-free level.
To examine the impact of backaction on the conjugate quadratures, we now also realize a tomography of the mechanical state under a pump-cavity BAE measurement through the probe cavity, by applying another set of four BAE tones
(\fref{fig:scheme}D). 
The tomography measures the variance of a generalized collective quadrature, e.g.~$X_+^{\phi}$. The tomographic angles (e.g.~$\phi$) for the probe-cavity BAE measurement are now determined by the pump-cavity BAE setup.

The variance of the generalized quadrature is $\langle (X_+^{\phi} )^2 \rangle = \langle X_+ ^2 \rangle \cos^2 \frac{\phi}{2} +  \langle P_- ^2 \rangle \sin^2 \frac{\phi}{2} +  \langle X_+ P_- \rangle \sin \phi$ including a cross-correlation term $\langle X_+ P_- \rangle$.
While $\langle X_+ P_- \rangle$ is not directly accessible, the modeling \CHANGE{(Supplementary Materials, I.B.2.)} \CHANGE{shows} that it is negligible.
The amplitude of the tomography tones is chosen much smaller than those \CHANGE{driving} the pump cavity, by $\simeq - 15\,\rm dB$: thus we can neglect QBA exerted by the probe cavity in comparison to that from the pump cavity. Similar calibrations as for the pump cavity allow for a readout of the \CHANGE{fluctuations} in all the collective quadratures (Figs.~S4,~S5). As seen from \fref{fig:qmfs}D, the tomography shows a reasonable agreement with predictions based on the assumption that the backaction consists of QBA only, without adjustable parameters. The additional heating of the conjugate quadratures is associated to a small
$n_c^T \simeq 0.23$.


Finally, we use the four-tone setup to create and detect quantum entanglement between the two mechanical oscillators \cite{Entanglement,Groblacher}. Entanglement for continuous variable states can be characterized by inequalities \CHANGE{involving} $\langle X_+ ^2 \rangle$ and $\langle P_- ^2 \rangle$
\cite{Tan1999Duan,Duan}, \vari{and it can be stabilized \cite{Entanglement}, transient \cite{Groblacher,Teufel2020entangle}, or conditional on a measurement record \cite{Polzik2020entangle}.} According to the Duan criterion \cite{Duan,Drummond2009epr}, two oscillators are entangled if 
$\langle X_+^2 \rangle + \langle P_-^2 \rangle < 1$. 
In an earlier work \cite{Entanglement}, two-tone BAE tomography \cite{WoolleyBAE} allowed access to the conjugate quadratures $X_+$ and $P_+$ only, and $P_-$ had to be inferred based on other information. Using four-tone driving and four-tone BAE tomography, we can access both $X_+$ and $P_-$ \CHANGE{directly}.

\begin{figure*}[t]
  \begin{center}
    {\includegraphics[width=0.95\textwidth]{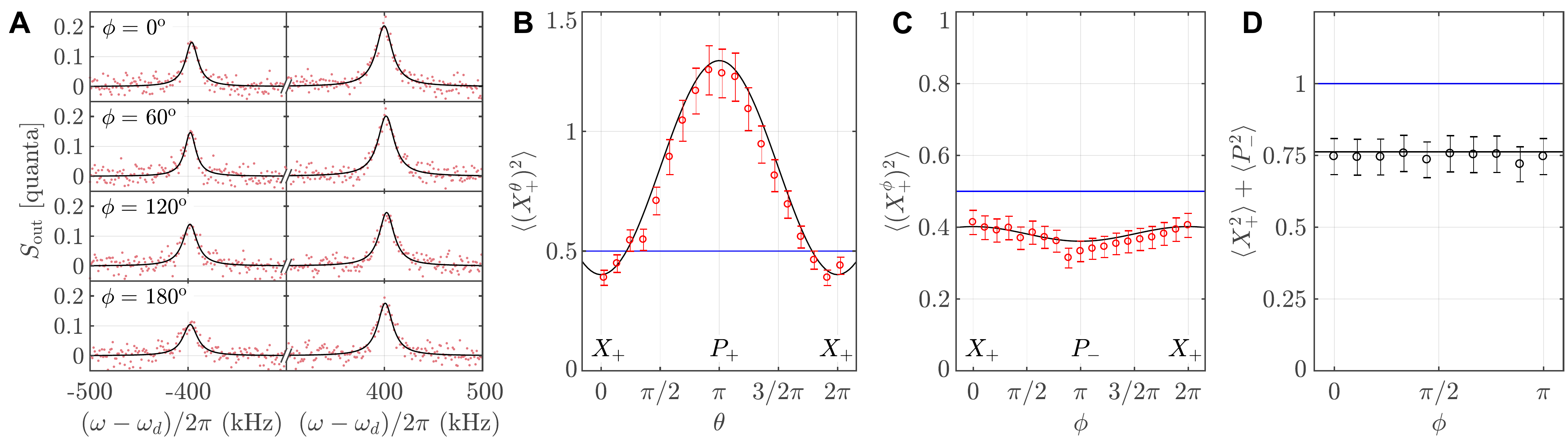}} 
    \caption{\textit{Direct measurement of \CHANGE{stabilized} quantum entanglement of two mechanical oscillators.}   \textbf{(A)} Output spectrum peaks in tomography carried out through the probe cavity while scanning between the noiseless quadratures $X_+$ and $P_-$. \CHANGE{At the different phase values $\phi = 0$ (top) to $\phi = \pi$ (bottom), the measurement probes the linear combinations: $\{X_+\}$, $\{ \frac{\sqrt{3}}{2}X_+ + \puoli P_-\}$, $\{ \puoli X_+ + \frac{\sqrt{3}}{2} P_- \}$, and $\{P_- \}$.} The solid lines are Lorentzian fits.  \textbf{(B)} Calibrated \CHANGE{variances} of the system of the noiseless quadrature  $X_+$ and the noisy $P_+$ swept by changing the probing phase. \textbf{(C)} Integrated variance from (A). 
    \textbf{(D)} Duan quantity extracted from (C) by summing up two generalized quadratures spaced by $\pi$. The horizontal axis is the starting phase in (C). The blue lines mark the vacuum fluctuation levels, and in (D), it gives the separability criterion of the quantum state. \CHANGE{The red-detuned effective coupling $|G_{j-}|/2\pi \simeq 121$ kHz, and $|G_{j+}/G_{j-}| \simeq 0.37$, corresponding to $C \simeq 490$. In (B), the detuning $\Omega/2\pi = 100$ kHz, and in (C-D),  $\Omega/2\pi = 400$ kHz. The black lines in (B-D) are theoretical predictions using $\langle X^2\rangle^T = 54$}.}
    \label{fig:entangle}
 \end{center}
\end{figure*}

In order to generate entanglement, we modify the BAE \CHANGE{setup} by reducing the \CHANGE{amplitudes of the blue tones} applied to the pump cavity \cite{Clerk13WangC,Meystre2013Sq,ClerkEnt2014}. \CHANGE{In this experiment we did not use cooling tones in the probe cavity and $\langle X ^2 \rangle^T = \langle X ^2 \rangle^{T,0}$.}
\CHANGE{The probe cavity is utilized for tomography, providing the spectra shown in \fref{fig:entangle}A whose integration gives the variances, as previously.}
\CHANGE{Scanning} the tomography between quadratures $X_+$ and $P_+$, we show in \fref{fig:entangle}B a large contrast between \CHANGE{$\langle P_+ ^2\rangle$ and $\langle X_+ ^2\rangle$, the latter reaching below the level of vacuum fluctuations}. \CHANGE{Changing the probe phase configuration to align} the probe QMFS with the pump QMFS $\{X_+, P_-\}$, we demonstrate in \fref{fig:entangle}C that the variances \CHANGE{inside this subspace are almost independent on the tomographic angle, as hinted by the modest variation
of peak amplitudes in \fref{fig:entangle}A, and that all measured variances remain clearly below the level of vacuum fluctuations.}
Finally, to assess the Duan criterion, we show in \fref{fig:entangle}D the sum of variances of two orthogonal quadratures  $X_+^\phi$ and $X_+^{\phi + \pi}$ taken for different values of $\phi$.
The sum satisfies the Duan criterion by $1.4\,\rm dB$ margin,
constituting a direct and robust measurement of the entanglement of two massive mechanical oscillators.

We have demonstrated the monitoring of two quadratures of an effective oscillator without quantum-backaction disturbance to the oscillator, which according to a common paradigm is not possible. This allows for a complete characterization of a weak classical force driving an oscillator and will have practical relevance when cavity optomechanical techniques will become available for sensitive measurements in the quantum regime at room temperature. The directly demonstrated entanglement of two massive oscillators
is a further tool to reduce intrinsic noise in such measurements (Fig.~S7). Combined with squeezing of probe electromagnetic fields and phase-sensitive amplification, it could allow for noiseless monitoring of weak external forces.

\renewcommand{\bibsection}{\subsection*{REFERENCES AND NOTES}}


\newpage

\begin{acknowledgments} We acknowledge the facilities and technical support of Otaniemi research infrastructure for Micro and Nanotechnologies (OtaNano) that is part of the European Microkelvin Platform. \textbf{Funding:} This work was supported by the Academy of Finland (contracts 308290, 307757, 312057), by the European Research Council (615755-CAVITYQPD), and by the Aalto Centre for Quantum Engineering. The work was performed as part of the Academy of Finland Centre of Excellence program (project 336810). We acknowledge funding from the European Union's Horizon 2020 research and innovation program under grant agreement No.~732894 (FETPRO HOT). M.J.W. acknowledges support from ARC Centre for Engineered Quantum Systems and AFOSR FA 2386-18-1-4026 \CHANGE{through the Asian Office of Aerospace Research and Development (AOARD)}. \textbf{Author contributions:} M.A.S.~supervised the project was involved in all subsequent stages. L.M.L.~and C.F.O.-K.~designed and fabricated the devices. L.M.L.~carried out most of the experiments. M.J.W.~suggested the experiment, and developed the theory. All authors participated in the writing of the paper. \textbf{Competing interests:}  The authors declare no competing interests. \textbf{Data and materials availability:} All data needed to evaluate the conclusions in the paper are present in the paper or the Supplementary Materials. 
\end{acknowledgments}

\end{document}